# Electronic Janus lattice and kagome-like bands in coloring-triangular MoTe$_2$ monolayers

Le Lei[1+], Jiaqi Dai[1+], Haoyu Dong [1+], Yanyan Geng[1], Feiyue Cao[1], Cong Wang[1], Rui Xu[1], Fei Pang[1], Fangsen Li[2,3], Zhihai Cheng[1,*], Guang Wang[4,5,*] and Wei Ji[1,*]

[1]*Beijing Key Laboratory of Optoelectronic Functional Materials & Micro-nano Devices, Department of Physics, Renmin University of China, Beijing 100872, People's Republic of China*

[2]*Vacuum Interconnected Nanotech Workstation, Suzhou Institute of Nano-Tech and Nano-Bionics, Chinese Academy of Sciences, Suzhou 215123, China*

[3]*School of Nano-Tech and Nano-Bionics, University of Science and Technology of China, Hefei 230026, China*

[4]*Department of Physics, College of Sciences, National University of Defense Technology, Changsha 410073, China*

[5]*State Key Laboratory of Low-Dimensional Quantum Physics, Department of Physics, Tsinghua University, Beijing 100084, China*

**Abstract:** Polymorphic structures of transition metal dichalcogenides (TMDs) host exotic electronic states, like charge density wave and superconductivity. However, the number of these structures is limited by crystal symmetries, which poses a challenge to achieving tailored lattices and properties both theoretically and experimentally. Here, we report a coloring-triangle (CT) latticed MoTe$_2$ monolayer, termed CT-MoTe$_2$, constructed by controllably introducing uniform and ordered mirror-twin-boundaries into a pristine monolayer in molecular beam epitaxy. Low-temperature scanning tunneling microscopy and spectroscopy (STM/STS) together with theoretical calculations reveal that the monolayer has an electronic Janus lattice, i.e., an energy-dependent atomic-lattice and a pseudo-Te sublattice, and shares the identical geometry with the Mo$_5$Te$_8$ layer. Dirac-like and flat electronic bands inherently existing in the CT lattice are identified by two broad and two prominent peaks in STS spectra, respectively, and verified with density-functional-theory calculations. Two types of intrinsic domain boundaries were observed, one of which the electronic-Janus-lattice feature maintains, implying potential applications as an energy-tunable electron-tunneling barrier in future functional devices.

[+]These authors contributed equally: Le Lei, Jiaqi Dai, Haoyu Dong
[*]Email: zhihaicheng@ruc.edu.cn, wangguang@nudt.edu.cn and wji@ruc.edu.cn

**Introduction**

Two-dimensional (2D) materials received increasing attention due to their exotic electronic and optical properties[1-12]. Recently, it comes into another intriguing property of 2D materials that engineering of their rich polymorphs showing diverse properties for wide applications[13-15]. Polymorph refers to the concept that a given composition with a variety of different crystal structures, including single-element materials and compounds. For instance, borophene possesses a highly polymorphic characteristic. It exhibits many atomic structures due to the complexity of bonding motifs[16], in which a series of exotic properties, including massless Dirac fermions[17] and 1D nearly free-electron states[18], were found. Additional states, such as superconductivity[19], charge density wave (CDW) [20-22], and nontrivial topological states[23] were recently found in layered transition metal dichalcogenides (TMDs), offering a particular platform to investigate fundamental condensed matter physics in the two-dimensional limit. Monolayer TMDs were successfully fabricated in many polymorphic phases, such as the hexagonal (1H), octahedral (1T) and monoclinic (distorted octahedral) (1T')[24] phases, showing phase-related properties. For example, the monolayer (ML) 1T-NbSe$_2$ exhibits a √13×√13 CDW order and a correlated magnetic insulating state[25], but its 1H counterpart possesses a 3×3 CDW order and superconductivity[19].

Mirror twin boundaries (MTBs)[26, 27] were demonstrate to be another strategy to introduce additional exotic electronic states in chalcogen-deficient 1H-MoS$_2$[28], -MoSe$_2$[26], and - MoTe$_2$[29] monolayers. These MTBs are metallic and show a high density of states (DOS)[30] near the Fermi level ($E_F$), which usually lead to the formation of charge orders, like Peierls-type CDW[31, 32] or Tomonaga-Luttinger liquid[33, 34] at low temperature. MTBs, in form of chalcogen-sharing lines, develop in three equivalent zigzag (ZZ) directions of the TMD monolayer lattice. This three-fold equivalence enables the MTBs to form triangular structures and dense networks[26], which could serve as block units for potentially building well-defined, like kagome[35, 36] or coloring-triangle[37] lattices. Although it poses a huge challenge to experimental realization, a TMD layer consisting of ordered and uniformly sized MTB triangles, namely an MTB-triangle lattice[38, 39], could be a TMD phase exhibiting a well-defined lattice symmetry. Therefore, this strategy allows the expansion of the family of polymorphic TMD phases which are essential for exploring exotic electronic states in the 2D limit.

In this work, we constructed a coloring-triangular (CT) lattice in a MoTe$_2$ (CT-MoTe$_2$) monolayer comprised of uniform-sized and orderly arranged MTB triangles and normal MoTe$_2$ domains embedded among MTBs. This CT-MoTe$_2$ monolayer was theoretically proposed and experimentally prepared using a controllable annealing process to an as-grown MoTe$_2$ monolayer. The geometric and electronic structures of CT-MoTe$_2$ were measured using scanning tunneling microscopy/spectroscopy (STM/STS) and verified with first-principles calculations, which reveal an electronic Janus lattice showing two energy-dependent lattices. Further STS measurements in CT-MoTe$_2$ show two prominent peaks near the Fermi level ($E_F$), which are related to two flat electronic bands inherently existing in CT lattices, as observed in our theoretical calculations. Furthermore, we found a domain boundary in CT-MoTe$_2$, which becomes invisible in certain energy windows. In other words, it behaves like an energy-tunable barrier for electrons flowing through. This work sheds considerable light on the identification of more complicated but uniform polymorphs of TMD monolayers which exhibit exotic electronic phenomena in 2D systems.

**Results**

Figure 1a shows a typical STM topographic image of the MoTe$_2$ sample after post-annealing of an epitaxially grown monolayer on a HOPG substrate at ~ 513K. The accompanying STM current image (Fig. 1b) clearly manifests the coexisting of 1H- and 1T′-MoTe$_2$ phases[24]. A substantial portion of the 1H regions is covered by high-density mirror-twin boundaries (MTBs), as shown in Fig. 1b and Supplementary Fig. 1, forming various dense MTB networks and/or triangles in different sizes. Figure 1c and 1d show representative high-resolution topography images of the MTB structures, which appear as single (Fig. 1c) and double (Fig. 1d) bright stripes for the empty and occupied states, respectively. The *dI/dV* spectra of 1H- and 1T′-MoTe$_2$, as shown in Fig. 1e and Supplementary Fig. 2d, reveal a semiconducting bandgap of ~1.9 eV, consistent with a previous value of 2.01 eV[32], and a semi-metallic gapless feature[40], respectively. A distinct narrow U-shaped gap around $E_F$ was observed in the *dI/dV* spectrum acquired on the MTB structure (Fig. 1f), where two sharp peaks residing at the two sides of the gap, ascribed to a Peierls-type CDW[31, 32] or Tomonaga-Luttinger liquid state [33].

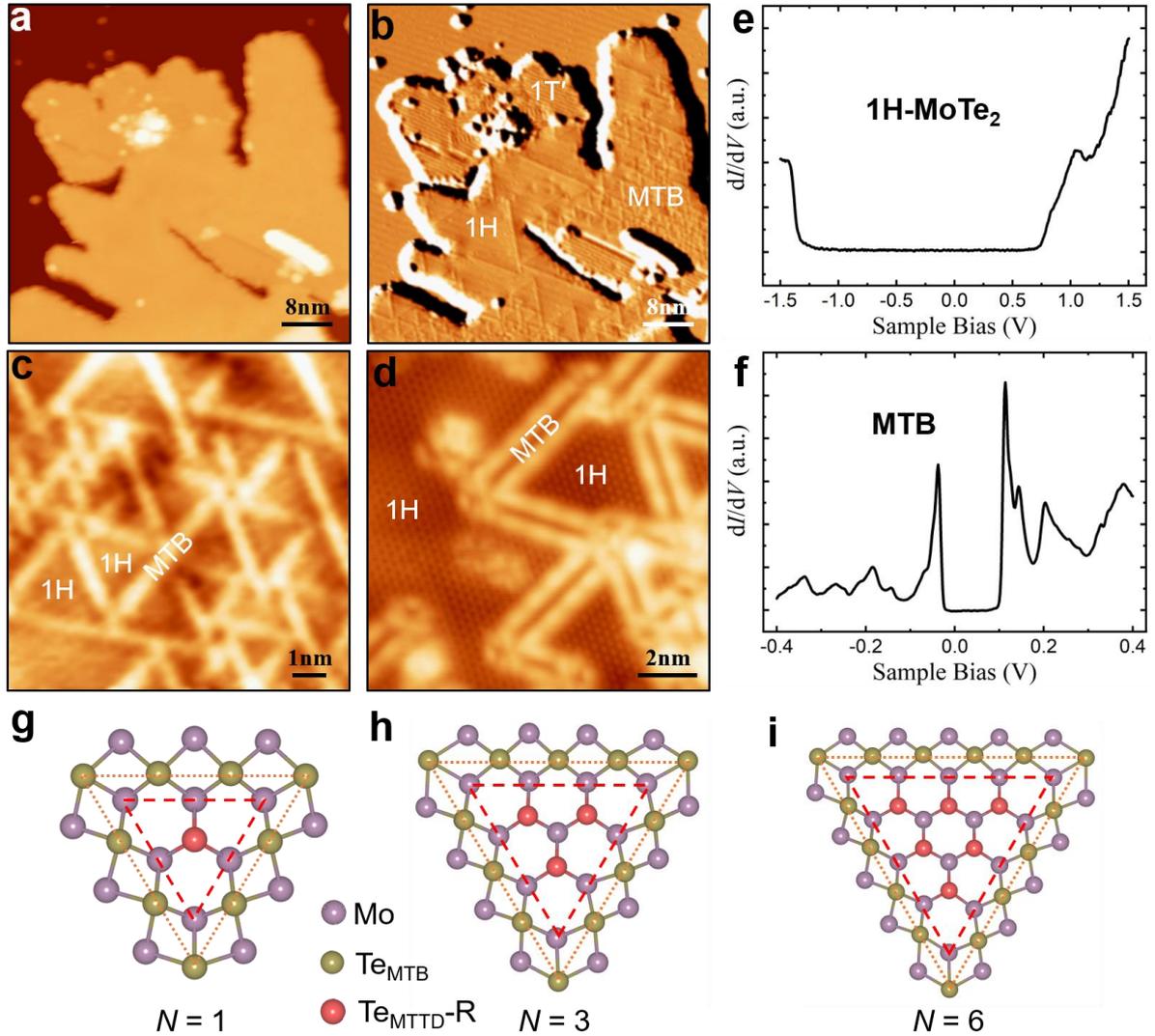

**Figure 1. Morphology and polymorphs of monolayer MoTe₂. a, b** Large-scale STM topography (**a**) and current (**b**) images of the synthetic MoTe$_2$. Local 1H- and 1T'-MoTe$_2$ phases, and mirror-twin boundaries (MTB) are labelled as 1H, 1T' and MTB, respectively. **c** Magnified STM topographic image showing 1H-MoTe$_2$ domains and MTB networks. **d** Atomically resolved STM topographic image of MTB triangles. **e, f** Typical $dI/dV$ spectra taken on the 1H-MoTe$_2$ domains (**e**) and MTB networks (**f**), respectively. **g-i** Illustration of the atomic models of MTB triangles of different sizes of N. The red and brown balls represent Te atoms, the violet balls represent Mo atoms. N represents the number of the Te$_2$ unit (red balls) in the 1H-MoTe$_2$ domain. Red hashed triangles outline Mo-terminated triangular domains (MTTD), while their associated mirror-twin-boundaries (MTB) are highlighted using orange dotted triangles (hereinafter). Scanning parameters are (a, b) $V$= 2.6V, $I$= 100pA, 52nm×52nm; (c) $V$= 2.0V, $I$= 70pA, 10nm×10nm; (d) $V$= -1.1V, $I$= -80pA, 12nm×12nm.

The interconnected MTBs of the MoTe$_2$ sample form triangular loop structures surrounding the pristine 1H-MoTe$_2$ domains (Fig. 1d) in which the sizes of the triangles are randomly distributed. Figures 1g–1i illustrate the atomic models of three MTB triangles in sizes of N=1, 3, and 6, where N represents the number of Te$_2$ units (red in Fig. 1g–1i) in the 1H-MoTe$_2$ domain. Here, an MTB triangle (highlighted by orange dotted triangles in Fig. 1g–1i) contains a Mo-terminated triangular domain (MTTD) of pristine 1H-MoTe$_2$ (highlighted by red hashed triangles in Fig. 1g–1i) and its surrounding Te lines at MTBs. The size of MTTD is governed by the chalcogen deficiency, as demonstrated in the literature[29]. It was thus an effective route to control the size of MTB triangles that controllably post-growth removal of chalcogen atoms, which was recently achieved by annealing the sample at a certain temperature during a certain period of time[26, 41, 42]. In the high MTB density limit, the smallest MTB triangle (N=1, Fig. 1g) dominantly presents, which, most likely, has a sufficiently large condition window for experimental realization and is expected to host exotic electronic states. We thus use *N*=1 MTB triangles (MTTDs) for illustration in our theoretical proposal.

Figure 2a depicts an ordered triangular lattice of the smallest MTTDs, highlighted using red dashed triangles, arranged in a corner-to-edge manner, while blue solid triangles displayed in Fig. 2b highlight a hexagonal lattice of the smallest MTTDs oppositely oriented to the red triangle and assembled in a corner-to-corner manner. Their interstitial regions are filled with the Te lines of MTBs (gray dotted lines in Fig. 2a and 2b). These MTTDs, together with the MTBs, form a novel polymorphic MoTe$_2$ phase, the lattice model of which was recently proposed in theory as the coloring-triangular (CT) lattice (Fig. 2c), hosting kagome-like electronic bands[37]. Thus, we denote this phase as the CT-MoTe$_2$ phase. The CT lattice is a variant of the kagome lattice which is an interesting and well-defined lattice. The bandstructures of the CT lattice consist of many sets of two Dirac bands and one flat band.

Density functional theory calculations were further carried out to elucidate the existence of the CT-MoTe$_2$ monolayer. Figure 2d shows the fully relaxed atomic structure of the CT-MoTe$_2$ monolayer, which is, as we proposed, comprised of oppositely oriented N=1 MTTDs (red and blue triangles) being separated by the Te lines of MTBs (the gray dotted triangle). Such an arrangement yields a triangular superlattice with a lattice constant of 12.66 Å (black dashed rhombuses in Fig. 2a–2c. However, those central Te atoms of the MTTDs, regardless of their orientations, spatially reside in a smaller triangular lattice with a lattice constant of 7.31 Å. We denote it as the Te pseudo-lattice in the CT-MoTe$_2$ monolayer that its periodicity is expected to show under certain selected energies. This smallest MTB monolayer shares the identical geometry and chemical ratio with the recently found Mo$_5$Te$_8$ monolayer[38].

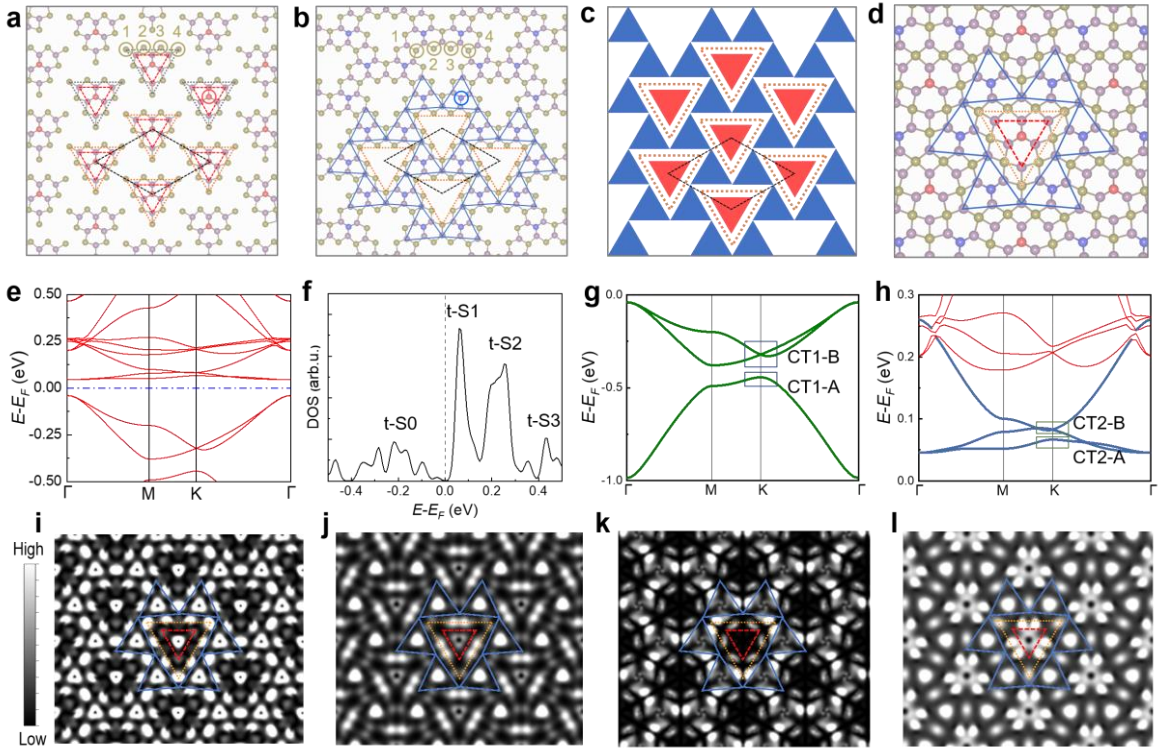

**Figure 2. Theoretical atomic and electronic structures of the CT-MoTe$_2$ phase. a** Illustration of the triangular lattice of the smallest MTTDs (highlighted by red dashed triangles) inside the associated MTB triangles (orange dotted triangles, hereinafter), which are arranged in a corner-to-edge manner. The black dashed rhombus indicates the supercell of the lattice (hereinafter) **b** Illustration of the hexagonal lattice of the smallest MTTDs (highlighted by blue triangles) among the associated MTBs, connected in a corner-to-corner manner. The central Te atoms in MTTDs are highlighted by the red (Te$_{MTTD}$-R) and blue (Te$_{MTTD}$-B) circles, respectively. **c** Schematic of the formed coloring-triangle (CT) lattice of CT-MoTe$_2$, composed of both triangular (**a**) and hexagonal (**b**) lattice of the smallest MTTDs. **d** Illustration of the atomic structure of the CT-MoTe$_2$ phase. **e** Theoretical band structures of the CT-MoTe$_2$ monolayer. **f** Corresponding total density of states (DOS) of the CT-MoTe$_2$ monolayer. **g, h** Zoomed-in band structures of two sets of the CT bands display in green lines (CT1) and blue lines (CT2). **i-l** 2D contour of visualized wavefunction norms at the K point (see blue rectangles in **g** and **h**) for CT1-A (**i**), CT1-B (**j**), CT2-A (**k**), and CT2-B (**l**), respectively. The isosurface values were kept fixed at 2×10$^{-4}$ e Bohr$^{-3}$.

Figure 2e shows the electronic band structure of the CT-MoTe$_2$ monolayer, in which a small energy bandgap opens at the Fermi level ($E_F$). An on-site Coulomb energy U=1.5 eV is mandatory to obtain the 0.09 eV bandgap, indicating correlated electronic characteristics of the CT-MoTe$_2$ monolayer. Two sets of flat-bands reside at approximately 0.05 and 0.20 eV (over $E_F$, hereinafter), respectively, which are denoted states t-S1 and t-S2 in the density of states (DOS) plotted in Fig. 2f. These two sets of flat bands, together with those Dirac-like dispersive bands below $E_F$ (t-S0 in Fig. 2g) and over those two flat-bands (t-S3 in Fig. 2g), are consistent with electronic features of the kagome-variant CT lattice. At least two sets (CT1 and CT2) of bands around $E_F$, primarily constituted of Te $p$ states and Mo $d$ states, show kagome-like features, as highlighted in green (CT1) and blue (CT2) in Fig. 2g and 2h. Each set contains a nearly flat-band and two highly dispersive bands showing Dirac-like behaviors around the K point, although a bandgap opens at K for the green set.

For the green set, we visualized their wavefunction norms of the -0.44 eV (CT1-A) state at the K point and the degenerated states above (-0.32 eV, CT1-B) in Fig. 2i and 2j, respectively. They, as highlighted using triangles, both show the pattern of the CT lattice. The CT1-A state is primarily distributed on the $p_z$ states of Te$_{MTTD}$-R and Te$_{MTB}$-1 and -4 atoms, while the CT1-B state is comprised of the bonding state of the $p_{xy}$ orbitals of Te$_{MTB}$-2 and -3 and the $p_{xy}$ state of Te$_{MTTD}$-B. By this means, CT1-A and CT1-B are energetically and spatially separated. The band structure (Fig. 2h) and visualized wavefunction norms (Fig. 2k and 2l) of set CT2 exhibit comparable patterns, namely a set of kagome-like bands and CT-symmetry appeared wavefunction norms. However, the contribution from Te$_{MTTD}$-R nearly eliminates in visualized wavefunction norm of the CT2-A state (0.07 eV) at the K point and those for the Te$_{MTB}$-1 and -4 atoms are from $p_{xy}$ orbitals (Fig. 2k). Unlike CT1-B, CT2-B is partially constituted of the *anti*-bonding state of the $p_{xy}$ orbitals of Te$_{MTB}$-2 and -3 (Fig. 2l), which explains why the CT2 set sits at higher energy and is thus unoccupied. Both pronounced electronic contributions from Te$_{MTTD}$-R and Te$_{MTTD}$-B in CT1-A (Fig. 2i), most likely, result in a smaller apparent lattice period of the surface, which is expected to be selectively visualized at certain energy windows. We thus term this feature as the Janus electronic states, as we elucidated in our following experiments.

We post-annealed our samples shown in Fig. 1 at an even higher temperature of 616 K to prepare the CT-MoTe$_2$ monolayer. Figure 3a shows a STM topographic image of the sample. Several wire-like features were observed at the edges of the monolayer islands, ascribed to Mo$_6$Te$_6$ nanowires (see Supplementary Fig. 4), clearly demonstrating a phase transition occurred at the edges under a chalcogen-deficient condition[43]. Moreover, those domains inside the islands show an ordered phase, distinctly different from the 1H- or 1T′- phase, as locally resolved and marked with white dashed lines in the STM current images (Fig. 3b and Supplementary Fig. 5). A magnified current image (Fig. 3c) more clearly displays the features of the emergent phase, which are consistent with the proposed CT-MoTe$_2$ phase. A structural model of the CT-MoTe$_2$ monolayer was replotted in Fig. 3d where the lattice vectors of the two Janus electronic lattices are displayed in red (atomic lattice) and blue (Te pseudo-lattice).

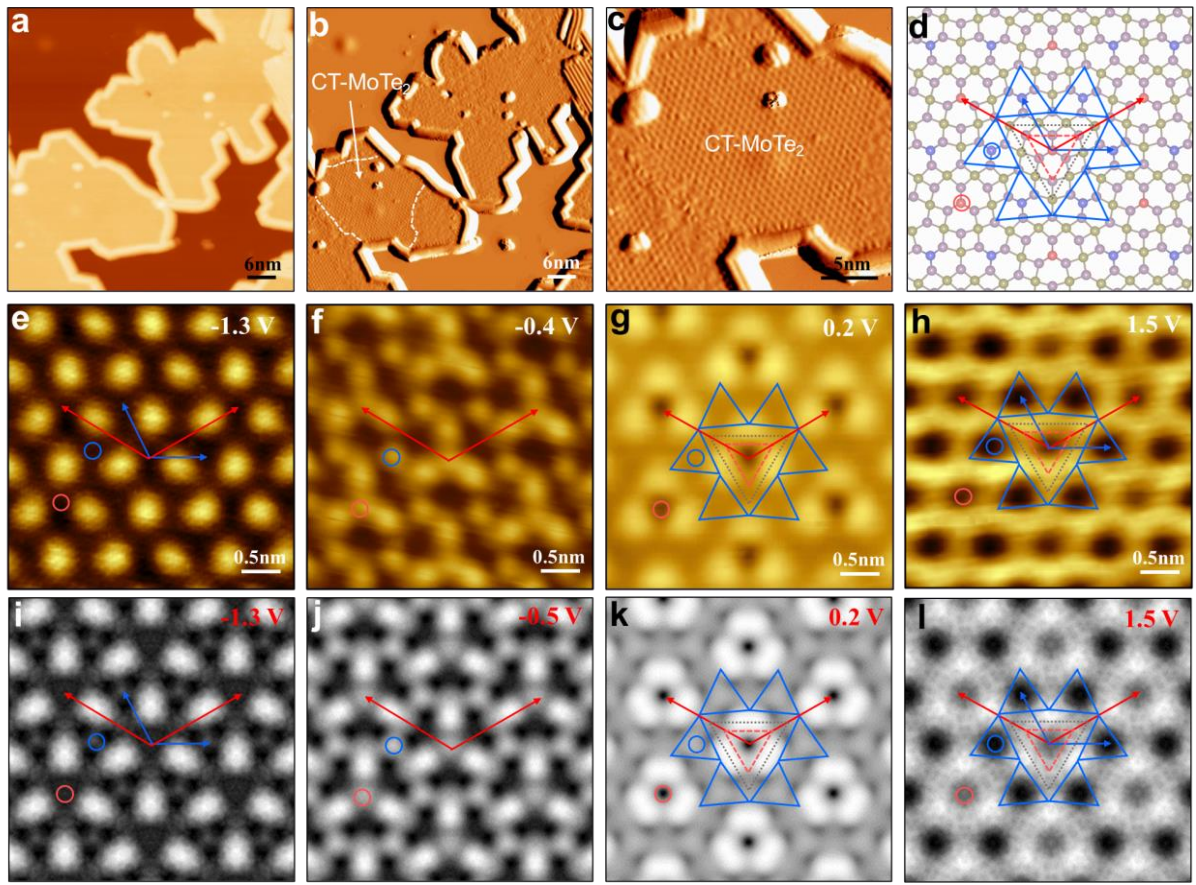

**Figure 3. Bias-dependent STM images of the CT-MoTe$_2$ phase. a, b** Large-scale STM topographic (**a**) and current (**b**) image of the post-annealed MoTe$_2$ monolayer. White dashed lines indicate the area of the formed CT-MoTe$_2$ phase. **c** Magnified STM current image of the CT-MoTe$_2$ area. **d** Structural model of the CT-MoTe$_2$ phase. **e-h** Bias-dependent STM topography images of the CT-MoTe$_2$ phase, showing an apparent electronic Janus lattice. Generally, the primitive atomic-lattice (red, largeer) and the pseudo-

sublattice (blue, smaller) are apparently observed within and out of the energy-range of (-1V, +1V), the lattice vectors of which were denotes using the red and blue arrows, respectively. **i-l** Simulated STM images of the CT-MoTe$_2$ phase. The red/blue and grey triangles highlight the MTTDs and MTB triangular segments of CT-MoTe$_2$. The Te$_{MTTD}$-R and Te$_{MTTD}$-B atoms are marked by the red and blue circles in (d-l), respectively. (a, b) $V$= -2.3V, $I$= -100pA; (c) $V$= -2.0V, $I$= -100pA.

Comparison of a series of bias-dependent experimental (Fig. 3e–3h) and theoretical (Fig. 3i–3l) STM images verifies that the prepared sample is the CT-MoTe$_2$ monolayer exhibiting the Janus electronic lattices. The experiment and theory are well consistent over a large range of bias voltages in terms of image appearance and apparent lattice periodicity. This series of images shows two apparent lattices, namely a larger one representing the atomic lattice (red in Fig. 3e–3h and a smaller one showing the electronic Te pseudo-sublattice (blue in Fig. 3e and 3h). The atomic lattice was exclusively imaged at -0.4 (Fig. 3f) and +0.2 V (Fig. 3g), showing apparent tri-spots features in the 12.2 Å lattice (indicated by the red vectors in Fig. 3e–3h). At higher bias voltages, namely -1.3 and +1.5 V, the pseudo-sublattice was imaged as bright spots (Fig. 3e) or black pits (Fig. 3h) with a lattice constant of 7.1 Å (indicated by the blue vectors). Both lattice constants are consistent with the theoretical values of 12.66 and 7.31 Å of the proposed CT-MoTe$_2$ model shown in Fig. 2d. It is also subtly noted that the bright spots in Fig. 3e show a certain chiral characteristic, and one kind of pits (representing Te$_{MTTD}$-R, denoted using the red circle) appears less dark than the other (Te$_{MTTD}$-B, the blue circle) in Fig. 3h. These features, consistent with those in the simulated images (Fig. 3i and 3l), verify again that the prepared sample is the CT-MoTe$_2$ monolayer and indicate the electronic nature of its pseudo-sublattice.A similar feature was observed in the Mo$_5$Te$_8$ layer which was, however, inclusively and tentatively assigned to CDW states[38, 39].

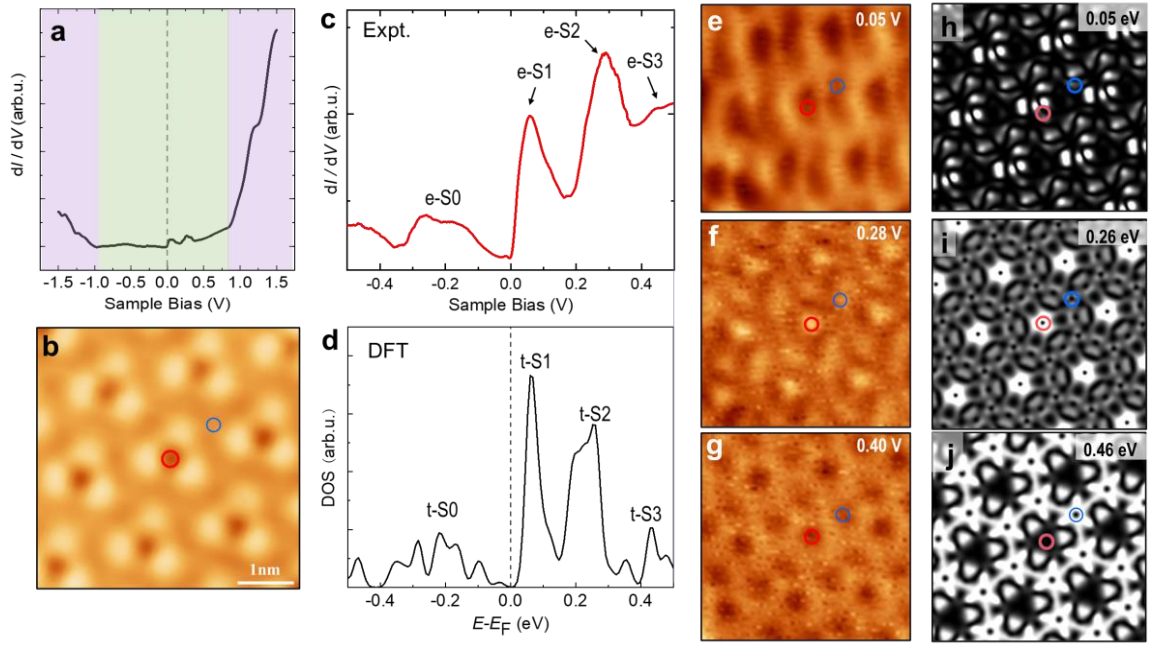

**Figure 4. STS measurements of the CT-MoTe$_2$ phase. a** Large-scale averaged $dI/dV$ spectrum of the CT-MoTe$_2$ phase, showing an apparent DOS gap (denoted using the green shadow) and appreciable in-gap states near the $E_F$. **b-d** STM topography image (**b**), magnified $dI/dV$ spectrum of the in-gap states (**c**) and total DOS of the CT-MoTe$_2$ phase. **e-j** Constant-current $dI/dV$ maps of (b) acquired at 0.05 V (**e**), 0.28 V (**f**) and 0.40 V (**g**), respectively, and their associated theoretically simulated maps derived from the wavefunction norms of the states sitting at 0.05 eV (**h**), 0.26 eV (**i**), and 0.48 eV (**j**) of the Γ point. The Te$_{MTTD}$-R and Te$_{MTTD}$-B atoms are marked by the red and blue circles in (b, d-l), respectively.

Tunneling spectra and d$I$/d$V$ maps of the CT-MoTe$_2$ monolayer (Fig. 4) verified, again, the experimental and theoretical consistency and visualized the appearance of the CT lattice. Within an apparent DOS gap (denoted using the green shadow in Fig. 4a) comparable to that of 1H-MoTe$_2$, appreciable in-gap states were observed near the $E_F$ (Fig. 4a). A magnified tunneling spectrum of the in-gap states was acquired and plotted in Fig. 4c. A pronounced dip at $E_F$ and two pronounced (e-S1 at 0.06 V and e-S2 at 0.28 V) and two wide peaks (e-S0 spanning from -0.35 – -0.03 V and e-S3 centered at 0.44 V) near the $E_F$. These four peaks were well reproduced in our DOS plot (Fig. 4d) as peaks t-S1 (0.06 eV), t-S2 (0.25 eV), t-S0 (from -0.39 – -0.01 eV), and t-S3 (centered at 0.43 eV). Peak t-S1 originates from state CT2-A, and state t-S2 is contributed from three less dispersive bands (CT3, Supplementary Fig. S3). Wide states t-S0 and t-S3 represent the dispersive bands of breathing kagome-like CT1-B (Fig. 2g and 2j) and CT4 (Supplementary Fig. S3), respectively. These well-consistent assessments indicate the existence of CT and kagome lattices and, at least, two sets of their associated flat bands (CT2-A and CT3) that may host strong electron-electron interaction.

Spatial maps of states e-S1, e-S2 and e-S3 were displayed in Fig. 4e–4g. Flat band e-S1 (Fig. 4e), showing a full width at half maximum of ~80meV, mostly records the spatial distribution of state CT2-A (Fig. 2k and 4h) which exhibits holes around atoms $Te_{MTTD}$-R and -B (red and blue circles in Fig. 4h, respectively). Highly localized charge density is clearly resolved for peak e-S2 on $Te_{MTTD}$-R (red circle in Fig. 4f), well consistent with the theoretical map of t-S2 (Fig. 4i). This map exhibits a lattice of ~12.66Å, which represents the unit cell of the atomic-lattice. The delocalized e-S3 (t-S3) is mostly distributed on the $Te_{MTB}$ atoms and both the $Te_{MTTD}$-R and -B atoms appear dark, as shown in Fig. 4g (4j), which exhibits an electronic lattice of ~7.31 Å (pseudo-sublattice). This interesting feature of the electronic Janus lattice in the CT-$MoTe_2$ monolayer was more comprehensively demonstrated in the energy-dependent STM/STS mappings shown in Supplementary Fig. 6.

**Discussion**

The CT-$MoTe_2$ monolayer, partially constituted of domain boundaries, also has domain boundaries where the electronic Janus lattice feature persists. Figure 5a and 5b show STM topography images of a domain boundary acquired at different bias voltages, which exhibits an inversion symmetry of atomic structures and is thus termed the IV boundary of the CT monolayer (DB-IV). DB-IV is characterized by a ZZ-arranged $Te_{MTTD}$-B atomic chain indicated by light and dark blue triangles and one inversion center was marked using a pink cross. In STM images, DB-IV is almost indistinguishable in Fig. 5a where the image exhibits the electronic pseudo-sublattice, while it is explicitly imaged in Fig. 5b, in which the atomic-lattice is displayed. These images indicate that the translation symmetry of the atomic-lattice breaks in DB-IV, as plotted in the atomic structure shown in Fig. 5c, but that of the electronic pseudo-sublattice is nearly maintained. Boundary DB-IV, separating two adjacent domains, behaves like an electronically tunable barrier resisting carrier flowing. If the CT-$MoTe_2$ was integrated into an electronic device, one would expect DB-IV could promote the gating efficiency at some certain gating voltages.

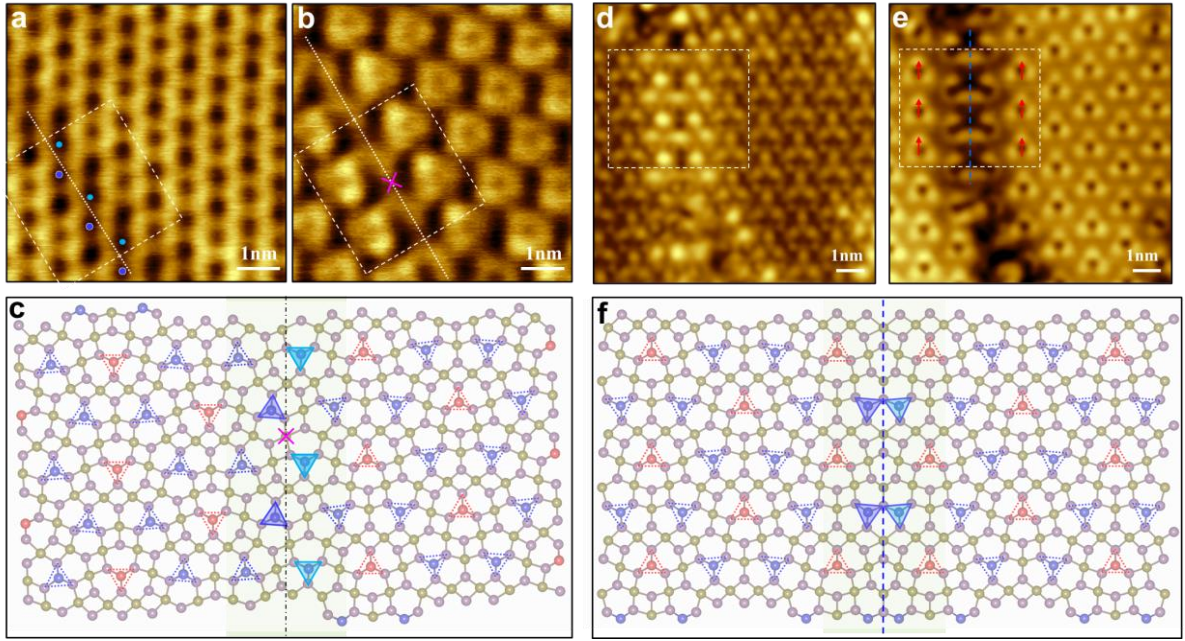

**Figure 5. Domain boundaries of the CT-MoTe$_2$ phase. a, b** STM topography images of the domain boundary with the inversion symmetry (named as DB-IV). The translation symmetry in the atomic-lattice (pseudo-sublattice) breaks (nearly preserves) across DB-IV, consistent with the found electronic Janus lattice of the CT-MoTe$_2$ phase. **c** Atomic structural model of DB-IV. An inversion-symmetric center at the domain boundary was marked by the pink cross. **d, e** STM topographic images of the domain boundary with the mirror twin symmetry (named as DB-MT). **f** Atomic structural model of DB-MT. Either atomic-lattice or pseudo-sublattice symmetry breaks across DB-MT. Te$_{MTTD}$-R and -B are marked by the small red and blue dashed triangles in (**c**) and (**f**). Dark- and light-blue shadowed triangles were used to denote Te$_{MTTD}$-B atoms at the boundaries to clearly demonstrate the symmetric features of DB-IV and -MT. (a) $V$= 1.34V, $I$= 80pA; (b) $V$=1.14V, $I$= 80pA; (d) $V$= -0.3V, $I$= -100pA; (e) $V$=0.3V, $I$= 100pA.

Another example of domain boundaries lies in two Te$_{MTTD}$-B atomic chains forming a mirror twin (MT) boundary. We thus term it as DB-MT, as shown in Fig. 5d and 5e, while Fig. 5f displays the corresponding atomic structural model of DB-MT. The translation symmetry of either the atomic-lattice or the pseudo-sublattice is broken across this boundary, which may lead to emerging properties subject to future experimental and theoretical studies. In addition, a domain boundary usually hosts distinct electronic states, which is possible to be applied in advanced functional devices. For example, the electronic "transparency" of the Janus DB-IV could be manipulated with an applied gate voltage, which could be potentially utilized in memory, logic operations and other devices.

**Conclusions**

In summary, we successfully synthesized the CT-MoTe$_2$ monolayer by introducing the highest MTB density orderly and uniformly into a pristine MoTe$_2$ monolayer using high-temperature post-growth annealing of MBE-grown MoTe$_2$ monolayers. In addition to flat electronic bands and Dirac-like states that the CT lattice symmetry inherently exhibit, the CT-MoTe$_2$ monolayer shows energy-dependent electronic Janus lattices, including the original atomic-lattice and an electronic Te pseudo-sublattice. Two types of domain boundaries were observed in the CT-MoTe$_2$ monolayer, one of which the electronic-Janus-lattice feature maintains implying application potentials in future functional devices. The atomic arrangement of CT-MoTe$_2$ inspires us to further expand the family of polymorphs in CDW phases of TMDs. A straightforward strategy lies in combining honeycomb-arranged CDW units centering an inverted CDW unit in a supercell, whose structural characteristic follows that of the CT-MoTe$_2$ (Supplementary Fig. 9). It was demonstrated that CDW structures in TMDs could be "condensed" selectively by point defects[20, 44], indicating the above scenario highly promising. Our work offers an effective route to artificially build structural polymorphs in TMDs that host exotic electronic properties to be explored.

**Methods**

**Sample preparation.** The single-layer MoTe$_2$ films were grown on highly oriented pyrolytic graphite (HOPG) substrate in a home-built MBE system with a base pressure of $3.0 \times 10^{-10}$ Torr. The highly oriented pyrolytic graphite (HOPG) substrate was freshly cleaved in air and immediately loaded into the ultra-high vacuum (UHV) chamber of MBE, then degassed at 773K overnight to remove contaminants. The high-purity Mo (99.999%) and Te (99.999%) were simultaneously evaporated from an electron beam evaporator and a Knudsen cell, respectively. The temperature of HOPG during growth was ~ 513K. After growth, all samples were followed by annealing with either a growth temperature maintained or higher temperature (616K).

**STM measurements.** The samples were transferred to another UHV chamber with LT-STM (PanScan Freedom, RHK) for the following STM measurements. All STM/STS measurements were performed at 9K with a chemically etched W tip calibrated on a clean Ag(111) surface (Supplementary Figure. 10). The STM images were acquired in constant-current mode. The $dI/dV$ spectra were obtained by using a standard lock-in amplifier with bias modulation ~ 5 mV at 857 Hz. All STM images were processed by Gwyddion and WSxM[45] software.

**DFT calculations.** Density functional theory calculations were performed using the generalized gradient approximation for the exchange-correlation potential, the projector augmented wave method, and a plane-wave basis set as implemented in the Vienna Ab initio Simulation Package (VASP)[46]. The energy cutoff for plane wave was set to 500 eV for invariant volume structural relaxation of freestanding CT-MoTe$_2$ monolayers. A dispersion correction was made at the van der Waals density functional (vdW-DF) level, with the optB86b functional for the exchange potential[47]. During all structural relaxations, all atoms were fully relaxed until the residual force per atom was less than $2\times10^{-4}$ eV Å$^{-1}$ and the energy convergence criteria was $1\times10^{-5}$ eV. A $7\times7\times1$ k-mesh was used to sample the first Brillouin zone in all calculations. An effective on-site Coulomb energy U=1.5 eV was considered in all calculations. The thickness of the vacuum layer is set to 15 Å. In plotting DOS spectra, a Gaussian smearing of 0.04 eV was used. The energy level of EF was set to energy zero in DOS and band structure calculations.

## Acknowledgments

This project is supported by the National Key R&D Program of China (Grant No. 2018YFE0202700), the National Natural Science Foundation of China (NSFC) (No. 21622304, 61674045, 11604063, 11974422, 12104504), Strategic Priority Research Program and Key Research Program of Frontier Sciences and Instrument Developing Project (Chinese Academy of Sciences, CAS) (No. No. XDB30000000, QYZDB-SSW-SYS031, No. YZ201418). Z. H. Cheng was supported by Distinguished Technical Talents Project and Youth Innovation Promotion Association CAS, the Fundamental Research Funds for the Central Universities and the Research Funds of Renmin University of China [No. 21XNLG27 (Z.C.), No. 22XNKJ30 (W.J.), No. 22XNH095 (H.D.)]. Calculations were performed at the Physics Lab of High-Performance Computing of Renmin University of China, Shanghai Supercomputer Center and Beijing Supercomputing Center.

## Author contributions

Z.C., G.W. and W.J. conceived the research project. L.L., H.D., Y.G. and Z.C. performed the STM experiments and analysis of STM data. R.X., F.P. and F.L. helped in the experiments. J.D., F.C., C.W. and W.J. performed the DFT calculations. L.L., J.D., F.C., Z.C. and W.J. wrote the manuscript with inputs from all authors.